\documentclass{inprep}
\usepackage{graphicx}

\newcommand{\Tr}{\mathop{\rm Tr}\nolimits}
\newcommand{\MS}{$\overline{\rm MS}$}
\newcommand{\ch}{\mathop{\rm ch}\nolimits}
\newcommand{\sh}{\mathop{\rm sh}\nolimits}
\renewcommand{\th}{\mathop{\rm th}\nolimits}
\newcommand{\tfrac}[2]{{\textstyle\frac{#1}{#2}}}

\begin{document}

\begin{center}
\Large Budker Institute of Nuclear Physics
\\[35mm]
\large A.~G.~Grozin, O.~I.~Yakovlev
\\[10mm]
\large Baryonic currents and their correlators
\\[3mm]
\large in the Heavy Quark Effective Theory
\\[20mm]
\Large Budker INP 92-59
\\
\vfill
NOVOSIBIRSK
\\[2mm]
1992
\end{center}
\thispagestyle{empty}
\newpage

\begin{center}
\bf Baryonic currents and their correlators
\\
\bf in the Heavy Quark Effective Theory
\\[3mm]
\it A.~G.~Grozin$^1$, O.~I.~Yakovlev$^2$
\\[3mm]
\rm Budker Institute of Nuclear Physics,
\\
630090 Novosibirsk 90, Russia
\\[5mm]
ABSTRACT
\end{center}
\begin{quotation}
HQET currents with the quantum numbers of the ground-state baryons
are discussed.
One-loop anomalous dimensions are calculated,
and exact one-loop matching to QCD currents is found.
Two-point correlators of these currents are calculated
taking into account $d\le9$ terms of the OPE.
Sum rules for heavy baryons $\Lambda_Q$ and $\Sigma^{(*)}_Q$ are analyzed.
Three-point correlators of two baryonic currents and a heavy-heavy
velocity changing current are calculated with the same accuracy.
The baryonic Isgur-Wise form factors are estimated
from the corresponding sum rules.
\end{quotation}
\vfill
\begin{flushright}
\copyright{} Budker Institute of Nuclear Physics
\end{flushright}
\addtocounter{footnote}{1}
\footnotetext{Internet address: \tt GROZIN@INP.NSK.SU}
\addtocounter{footnote}{1}
\footnotetext{Internet address: \tt O\_YAKOVLEV@INP.NSK.SU}
\thispagestyle{empty}
\newpage

{
\parindent=0pt
\begin{tabular}{r}
\hline
\hspace{112mm}
\end{tabular}
\par
}

\vspace{50mm}

\section{Introduction}
\label{Intro} \setcounter{equation}{0}

It was noticed long ago~\cite{Shuryak} that properties of hadrons
with a single heavy quark are simple: it's mass (and hence flavor)
and spin orientation are irrelevant to the leading order in $1/m$.
Recently this qualitative physical picture was incorporated into the formal
framework of the Heavy Quark effective Theory (HQET)~\cite{EH,GG}.
The heavy quark spin symmetry~\cite{IW} leads to relations
among meson~\cite{IW,FGGW} and baryon~[6--8] 
form factors.
An elegant general discussion can be found in~\cite{Falk}.
Not only the orientation but also the magnitude
of the heavy quark spin is irrelevant,
so this symmetry is extended to the superflavor symmetry~\cite{super}.

Nonperturbative methods (such as sum rules~\cite{SVZ})
are needed to obtain hadron properties in HQET.
HQET sum rules for mesons and baryons were first considered in~\cite{Shuryak}.
They should be improved in order to take into account
the currents' renormalization in HQET.
One-loop anomalous dimensions of HQET meson currents were found in~\cite{SV},
and two-loop ones---in~\cite{BG}.
HQET meson sum rules with the proper account of these effects
were considered in~\cite{BG2}.
the meson Isgur-Wise form factors were considered~\cite{R}
in the framework of three-point sum rules~\cite{ISNR}.

Baryonic currents and sum rules in QCD were considered
in a number of papers~[17--20]. 
One-loop anomalous dimensions of QCD baryonic currents
were found in~\cite{Peskin}, and two-loop ones---in~\cite{PS}
(where the perturbative correction to the sum rules was also obtained).
Three-point sum rules for baryon form factors were discussed in~\cite{NR}.

In the present work we discuss the HQET currents
with the quantum numbers of ground-state baryons.
In Sec.~\ref{SCur} we calculate their one-loop anomalous dimensions
(as was done for mesons in~\cite{SV}),
and obtain the exact one-loop matching of these HQET currents to QCD ones
(as was done for mesons in~\cite{EH}).
Note that non-logarithmic terms in the one-loop matching are useless
unless the two-loop anomalous dimensions are known.
They can be found using the methods of~\cite{BG};
we hope to return to this question later.

Our further analysis of the heavy baryons
is similar to the detailed light baryon analysis~\cite{BI}.
In Sec.~\ref{Stwo} we consider two-point correlators
of the HQET baryonic currents.
OPE for diagonal correlators contains even-dimensional terms;
perturbative terms ($d=0$), $\left<G^2\right>$ corrections ($d=4$),
$\left<\overline{q}q\right>^2$ terms ($d=6$),
and corrections to it ($d=8\ldots$) are taken into account.
OPE for nondiagonal correlators contains odd-dimensional terms
starting from $\left<\overline{q}q\right>$ ($d=3$);
$m_0^2$ and $\left<G^2\right>$ corrections ($d=5$ and $7$),
and $\left<\overline{q}q\right>^3$ terms ($d=9$) are also taken into account.
After that, the sum rules for $\Lambda_Q$ and $\Sigma^{(*)}_Q$
following from these correlators are analyzed.
The diagonal sum rules have large continuum contributions
because the spectral density grows like $\omega^5$.
In the nondiagonal sum rules it grows like $\omega^2$,
and continuum contributions are not so important.
But higher power corrections $d\ge7$) are poorly known.
Nevertheless, both types of sum rules give good results
consistent with each other.
We disagree with some results of~\cite{Shuryak}.

In Sec.~\ref{Sthree} we consider three-point correlators
of two HQET baryonic currents and a heavy-heavy velocity changing current.
We calculate diagonal and nondiagonal correlators
with the same accuracy as in the two-point case,
and obtain the baryon Isgur-Wise form factors from the sum rules.

\section{Baryonic currents in HQET}
\label{SCur} \setcounter{equation}{0}

Hadrons in HQET are classified according to the light fields'
angular momentum and parity $j^\pi$.
For the ground-state baryons, light quark spins
can add giving $j^\pi=0^+$ or $1^+$.
In the first case their spin wave function is antisymmetric;
Fermi statistics and antisymmetry in color
require antisymmetric flavor wave functions.
Hence light quarks must be different;
if they are $u$, $d$ then their isospin $I=0$.
In the second case the flavor wave function is symmetric;
if light quarks are $u$, $d$ then their isospin $I=1$.
This gives us the $\frac12^+$ baryon with $I=0$ called $\Lambda_Q$,
and the degenerate $\frac12^+$ and $\frac32^+$ baryons with $I=1$
called $\Sigma_Q$ and $\Sigma^*_Q$.

Baryonic currents have the form $\widetilde\jmath
=\varepsilon^{abc}(q_1^{Ta}C\Gamma\tau q_2^b)\Gamma'\widetilde{Q}^c$,
where $C$ is the charge conjugation matrix, $q^T$ means $q$ transposed,
$\tau$ is a flavor matrix
(symmetric for $\Sigma_Q$ and antisymmetric for $\Lambda_Q$),
and $\widetilde{Q}$ is the effective field
satisfying $\gamma_0\widetilde{Q}=\widetilde{Q}$.
We shall abbreviate it to
$\widetilde\jmath=(q^T C\Gamma q)\Gamma'\widetilde{Q}$.
A light quark pair with $j^\pi=0^+$ corresponds to the current
$a=q^T C\gamma_5 q$, and with $j^\pi=1^+$---to $\vec{a}=q^T C\vec{\gamma}q$.
One can easily check it using the $P$-conjugation $q\to\gamma_0 q$.
It is also possible to insert $\gamma_0$ into these currents
without changing their quantum numbers.
The current $\widetilde\jmath=a\widetilde{Q}$ has spin 1/2;
the current $\vec{\widetilde\jmath}=\vec{a}\widetilde{Q}$
contains spin 1/2 and spin 3/2 components.
The part $\vec{\widetilde\jmath}_{3/2}=\vec{\widetilde\jmath}
+\frac13\vec{\gamma}\vec{\gamma}\cdot\vec{\widetilde\jmath}$
satisfies the condition $\vec{\gamma}\cdot\vec{\widetilde\jmath}=0$
and hence has spin 3/2.
The other part $\vec{\widetilde\jmath}_{1/2}
=-\frac13\vec{\gamma}\vec{\gamma}\cdot\vec{\widetilde\jmath}
=\frac13\vec{\gamma}\gamma_5\widetilde{\jmath}_{1/2}$,
$\widetilde{\jmath}_{1/2}=\vec{a}\cdot\vec{\gamma}\gamma_5\widetilde{Q}$
has spin 1/2.
Finally we arrive at the currents
\begin{eqnarray}
&&\widetilde{\jmath}_{\Lambda1} = (q^T C\gamma_5 q)\widetilde{Q},
\quad
\widetilde{\jmath}_{\Lambda2} = (q^T C\gamma_0\gamma_5 q)\widetilde{Q},
\nonumber\\
&&\widetilde{\jmath}_{\Sigma1} =
(q^T C\vec{\gamma}q)\cdot\vec{\gamma}\gamma_5\widetilde{Q},
\quad
\widetilde{\jmath}_{\Sigma2} =
(q^T C\gamma_0\vec{\gamma}q)\cdot\vec{\gamma}\gamma_5\widetilde{Q},
\label{cur}\\
&&\vec{\widetilde\jmath}_{\Sigma^*1} = (q^T C\vec{\gamma}q)\widetilde{Q} +
\tfrac13\vec{\gamma}(q^T C\vec{\gamma}q)\cdot\vec{\gamma}\widetilde{Q},
\nonumber\\
&&\vec{\widetilde\jmath}_{\Sigma^*2} =
(q^T C\gamma_0\vec{\gamma}q)\widetilde{Q} +
\tfrac13\vec{\gamma}(q^T C\gamma_0\vec{\gamma}q)\cdot\vec{\gamma}\widetilde{Q}.
\nonumber
\end{eqnarray}
This classification follows ideas of~\cite{Falk,MRR};
the currents for $\Lambda_Q$ and $\Sigma_Q$ first appeared in~\cite{Shuryak}.

\begin{figure}[ht]
\begin{center}
\includegraphics{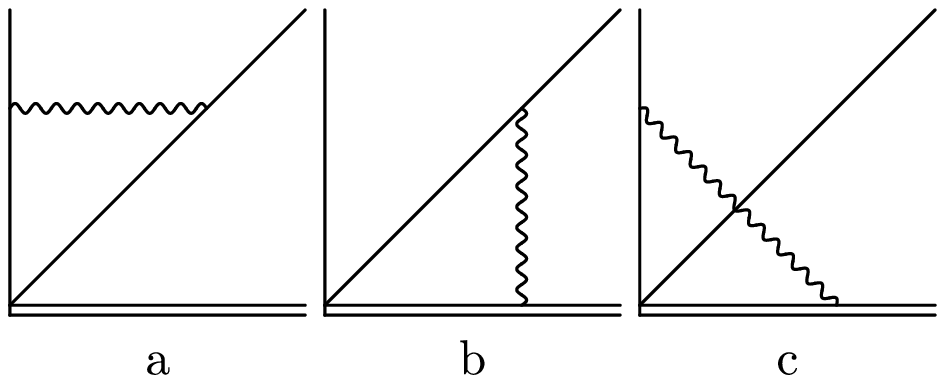}
\end{center}
\caption{One-loop proper vertex}
\label{F1}
\end{figure}

Now we consider the one-loop renormalization of these HQET currents.
In the \MS{} scheme (the space dimension $D=4-2\varepsilon$)
the bare fields are related to the renormalized ones by
$q_0=\bar{\mu}^{-\varepsilon}Z_q^{1/2}q$,
$\widetilde{Q}_0=\bar{\mu}^{-\varepsilon}\widetilde{Z}_Q^{1/2}\widetilde{Q}$,
where $\bar{\mu}^2=\frac{\mu^2 e^\gamma}{4\pi}$,
$\mu$ is the normalization point.
In the Feynman gauge the renormalization constants are
$Z_q=1-C_F\frac{\alpha_s}{4\pi\varepsilon}$,
$\widetilde{Z}_Q=1+2C_F\frac{\alpha_s}{4\pi\varepsilon}$,
where $C_F=\frac{N^2-1}{2N}$, $N=3$ is the number of colors.
The bare current $\widetilde{\jmath}_0=q_0 q_0 \widetilde{Q}_0
=\bar{\mu}^{-3\varepsilon}\widetilde{Z}_j\widetilde\jmath$,
or $qq\widetilde{Q}=\widetilde{Z}_\Gamma\widetilde\jmath$,
where $\widetilde{Z}_j=Z_q \widetilde{Z}_Q^{1/2}\widetilde{Z}_\Gamma$.
Then the matrix element
$\widetilde\Gamma=\left<0|qq\widetilde{Q}|qq\widetilde{Q}\right>
=\widetilde{Z}_\Gamma\left<0|\widetilde\jmath|qq\widetilde{Q}\right>$,
where the matrix element of $\widetilde\jmath$ is finite.
Calculating the ultraviolet divergence of the vertex $\Gamma$ Fig.~\ref{F1}
in the Feynman gauge, we obtain $\widetilde{Z}_\Gamma
=1+C_B\frac{\alpha_s}{4\pi\varepsilon}\left(\frac{H^2}4+2\right)$,
where $\gamma_\mu\Gamma\gamma_\mu=H\Gamma$,
$C_B=\frac{N+1}{2N}$\footnote{of course, it is a pure formality to write $N$
in formulae based on the fact that a baryon contains three quarks.}.
Using $Z_q$ and $\widetilde{Z}_Q$ we find
$\widetilde{Z}_j=\widetilde{Z}_\Gamma$;
hence the anomalous dimensions
$\widetilde{\gamma}_j=\frac{d\log\widetilde{Z}_j}{d\log\mu}$ are
\begin{eqnarray}
&&\widetilde{\gamma}_j = -C_B\frac{\alpha_s}{2\pi}\left(\frac{H^2}4+2\right),
\label{adim}\\
&&\widetilde{\gamma}_{\Lambda1} = -3C_B\frac{\alpha_s}{\pi},
\quad
\widetilde{\gamma}_{\Lambda2} = \widetilde{\gamma}_{\Sigma1}
= -\frac32 C_B\frac{\alpha_s}{\pi},
\quad
\widetilde{\gamma}_{\Sigma2} = -C_B\frac{\alpha_s}{\pi}.
\nonumber
\end{eqnarray}
They depend only on the light quark part of the current.

Now we are going to establish relation
between these HQET currents and QCD ones.
A QCD current $(q^T C\Gamma q)\Gamma'Q$ matches the corresponding effective
current $A\widetilde\jmath$ if they give the same physical matrix elements.
In order to calculate on-shell matrix elements, we have to use the on-shell
renormalization scheme in which propagators in the on-shell limit are free.
For the ``massless'' fields $q$, $\widetilde{Q}$,
the bare on-shell propagators get no corrections because loop integrals
are no-scale (ultraviolet and infrared divergences cancel).
Therefore the on-shell renormalized fields coincide with the bare ones:
$q=Z_q^{-1/2}q_{\rm os}$,
$\widetilde{Q}=\widetilde{Z}_Q^{-1/2}\widetilde{Q}_{\rm os}$.
Note however that although the expressions for renormalization constants
$Z_q$, $\widetilde{Z}_Q$ are the same, all divergences in them are infrared
ones because these $Z$ factors relate renormalized (ultraviolet-finite) fields.
For the massive quark field $Q$ we have $Q=Z_Q^{-1/2}Q_{\rm os}$,
$Z_Q=1+C_F\frac{\alpha_s}{4\pi}\left(\frac{2}{\varepsilon}-3L+4\right)$,
where $L=\log\frac{m^2}{\mu^2}$.
Note that the infrared divergence of the on-shell massive quark propagator
$Z_Q$ is the same as that of the effective quark propagator $\widetilde{Z}_Q$.

\begin{sloppypar}
We have the currents
$j=Z_\Gamma^{-1}Z_q^{-1}Z_Q^{-1/2}q_{\rm os}q_{\rm os}Q_{\rm os}$,
$\widetilde\jmath=\widetilde{Z}_\Gamma^{-1}Z_q^{-1}\widetilde{Z}_Q^{-1/2}
\allowbreak q_{\rm os}q_{\rm os}\widetilde{Q}_{\rm os}$.
Hence the on-shell matrix elements are
$\left<0|j|qqQ\right>=Z_\Gamma^{-1}Z_q^{-1}Z_Q^{-1/2}\Gamma$,
$\left<0|\widetilde\jmath|qqQ\right>=
\widetilde{Z}_\Gamma^{-1}Z_q^{-1}\widetilde{Z}_Q^{-1/2}\widetilde{\Gamma}$,
where the proper vertices $\Gamma$, $\widetilde{\Gamma}$ are depicted
on Fig.~\ref{F1}.
Hence we obtain the matching constant
$$
A = A_Q \frac{\Gamma/Z_\Gamma}{\widetilde{\Gamma}/\widetilde{Z}_\Gamma},
\quad
A_Q = \left(\frac{\widetilde{Z}_Q}{Z_Q}\right)^{1/2} =
1 + C_F\frac{\alpha_s}{4\pi}\left(\frac{3}{2}L-2\right).
$$
Here ultraviolet divergences cancel in $\Gamma/Z_\Gamma$,
$\widetilde{\Gamma}/\widetilde{Z}_\Gamma$ by definition;
infrared divergences cancel between these two expressions
because the infrared behavior of QCD and HQET is identical;
$A_Q$ is finite.
We choose all quark momenta in HQET to be zero;
this corresponds to the heavy quark momentum $mv$ ($v=(1,\vec{0})$) in QCD.
Then the HQET loops (Fig.~\ref{F1}) vanish.
The diagram Fig.~\ref{F1}a vanishes in $\Gamma$ (but not in $Z_\Gamma$).
Logarithmic terms in the matching conditions are determined
by the difference between QCD and HQET anomalous dimensions.
\end{sloppypar}

The choice of QCD currents corresponding to the HQET currents~(\ref{cur})
is not unique.
We restrict ourselves here to the simplest variant with
$\Gamma'\to\frac{1+\gamma_0}{2}\Gamma'\frac{1+\gamma_0}{2}$.
If we denote
$\left<(q^T C[\sigma_{\mu\nu},\Gamma]q)\frac{1+\gamma_0}{2}\Gamma'
\frac{1+\gamma_0}{2}\sigma_{\mu\nu}Q\right>=f(D)\left<j\right>$,
then the matching constant
$A=A_Q\left[1+C_B\frac{\alpha_s}{4\pi}
\left(\frac{f}{4}(L-3)+\frac12\frac{df}{dD}\right)\right]$.
Indices $\mu$, $\nu$ are purely space-like, hence the matching conditions
are the same for pairs of currents with $\Gamma\to\gamma_0\Gamma$.
Finally we obtain
(in the scheme where $\gamma_5$ anticommutes with all $\gamma_\mu$)
\begin{eqnarray}
&&(q^T C\gamma_5 q)\frac{1+\gamma_0}{2}Q = A_Q \widetilde{\jmath}_{\Lambda1},
\nonumber\\
&&-(q^T C\gamma_\mu q)
\frac{1+\gamma_0}{2}\gamma_\mu\gamma_5\frac{1+\gamma_0}{2}Q =
A_Q \left[1-C_B\frac{\alpha_s}{4\pi}(2L-4)\right]\widehat{\jmath}_{\Sigma1},
\label{match}\\
&&\frac{1}{D-1}(q^T C\gamma_\nu q)\frac{1+\gamma_0}{2}[(D-2)
(g_{\mu\nu}-v_\mu v_\nu)-\sigma_{\mu\nu}]\frac{1+\gamma_0}{2}Q
\nonumber\\
&&\quad{} = A_Q \left[1-C_B\frac{\alpha_s}{4\pi}(L-3)\right]
\widetilde{\jmath}_{\Sigma^*1\mu}
\nonumber
\end{eqnarray}
These equations should be used to obtain the QCD matrix elements from the HQET
ones calculated, for example, from the HQET sum rules (Sect.~\ref{Stwo}).

\section{Two-point correlators}
\label{Stwo}

Two-point correlators have the general structure
($\overline{\widetilde{\jmath}}=
(\overline{q}\overline{\Gamma}\tau^+C^{-1}\overline{q}^T
\overline{\widetilde{Q}}\overline{\Gamma}'$)
\begin{equation}
i\left<T\widetilde{\jmath}_1(x)\overline{\widetilde{\jmath}}_2(0)\right> =
\left(\Gamma'_1\frac{1+\gamma_0}{2}\overline{\Gamma}'_2\right)
\delta(\vec{x}) 2 \Tr \tau\tau^+ T \Pi(x_0).
\label{struct}
\end{equation}
The expression~(\ref{struct}) without the first factor
is the correlator in which the heavy quark spin is switched off
(its propagator is $\widetilde{S}(x_0)=-i\vartheta(x_0)$).
For both $j^\pi=0^+$ and $1^+$ there are two diagonal correlators
and one nondiagonal one (see~(\ref{cur})).
Correlators for the physical spin symmetry multiplets ($\Lambda_Q$,
$\Sigma_Q$, $\Sigma^*_Q$) are expressed via these ones~(\ref{struct})
due to the superflavor symmetry~\cite{super}.
The normalizing factor is $T=\frac14\Tr\Gamma_1\overline{\Gamma}_2$
for diagonal correlators and $T=\frac14\Tr\gamma_0\Gamma_1\overline{\Gamma}_2$
for nondiagonal ones.

\begin{figure}[p]
\begin{center}
\includegraphics{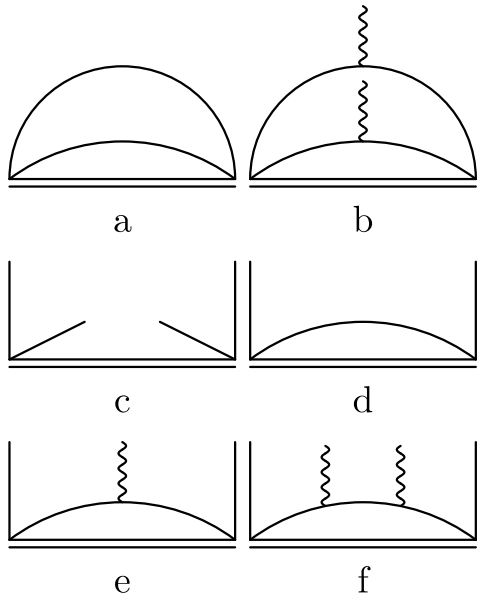}
\end{center}
\caption{Two-point correlator}
\label{F2}
\end{figure}

OPE for diagonal correlators contains even-dimensional terms.
We take into account the perturbative ($d=0$) term (Fig.~\ref{F2}a),
the gluon condensate ($d=4$) term (Fig.~\ref{F2}b),
the $\left<\overline{q}q\right>^2$ ($d=6$, 8\dots) terms (Fig.~\ref{F2}c).
The diagram Fig.~\ref{F2}d gives a $d=6$ contribution; it is of the order
of unknown perturbative corrections to the diagram Fig.~\ref{F2}c,
and is not taken into account.
We use the fixed-point gauge $x_\mu A_\mu(x)=0$ in which the heavy quark
does not interact with gluons.
The methods of calculation of correlators in this gauge
are reviewed in~\cite{TechRev}.
We obtain
\begin{eqnarray}
&&\Pi_{a+b}(t) = -\frac{N!\,\widetilde{S}(t)}{\pi^4 t^6}
\left[1 + c\frac{\pi\alpha_s\left<G^2\right>t^4}{32N(N-1)}\right],
\nonumber\\
&&\Pi_c(t) = \frac{N!\,\widetilde{S}(t)}{4N^2}
\left<\overline{q}(0)q(t)\right>^2.
\label{diag}
\end{eqnarray}
The terms $\Pi_{a+b}$ appear to be the same
for $\widetilde{\jmath}_1\widetilde{\jmath}_1$
and $\widetilde{\jmath}_2\widetilde{\jmath}_2$ correlators;
$c=1$ for $\Lambda_Q$ and $-\frac13$ for $\Sigma^{(*)}_Q$.
We have factorized a four-quark condensate in $\Pi_c$ into two two-quark ones.
In this approximation $\Pi_c$ is also the same in two diagonal correlators
(though the factorization approximation~\cite{SVZ} is thoroughly checked
only for products of vector and axial currents).
These correlators can't strictly coincide at least because
$\widetilde{\jmath}_1$ and $\widetilde{\jmath}_2$ have different
anomalous dimensions~(\ref{adim}).

The nonlocal quark condensate $\left<\overline{q}(0)q(x)\right>=
\left<\overline{q}q\right>\Big(1+\frac{m_0^2x^2}{16}+
\frac{\pi\alpha_s\left<G^2\right>}{96N}+\break\cdots\Big)$,
where in the last term the factorization is used
for $d=7$ quark-gluon condensates.
The Gaussian anzatz $\left<\overline{q}(0)q(x)\right>=
\left<\overline{q}q\right>\exp\left(\frac{m_0^2x^2}{16}\right)$
was proposed in~\cite{MR} instead.
The $x^4$ term in it is about 3 times larger
than in the factorization estimate.

The perturbative ($d=0$) and $\left<\overline{q}q\right>^2$ ($d=6$) terms
were first considered in~\cite{Shuryak}.
We agree with the perturbative result.
The $\left<\overline{q}q\right>^2$ term in~\cite{Shuryak} had different signs
for $\Lambda_Q$ and $\Sigma_Q$ while these signs are the same in~(\ref{diag}).
Because of this, the results~\cite{Shuryak} on $\Sigma_Q$
and its difference from $\Lambda_Q$ are incorrect.

OPE for nondiagonal correlators contains odd-dimensional terms.
The diagram Fig.~\ref{F2}d gives $d=3$, 5, 7\dots{} terms;
Fig.~\ref{F2}f---$d=7$\dots{} ones.
The diagram Fig.~\ref{F2}c contributes a $d=9$
$\left<\overline{q}q\right>^3$ term.
We obtain
\begin{eqnarray}
&&\Pi_d(t) =
-i\frac{N!\,\left<\overline{q}q\right>\widetilde{S}(t)}{\pi^2 Nt^3}
\left[1 + \frac{m_0^2 t^2}{16} + \frac{\pi\alpha_s\left<G^2\right>t^4}{96N}
\right],
\nonumber\\
&&\Pi_e(t) =
ic\frac{N!\,\left<\overline{q}q\right>\widetilde{S}(t)}{16\pi^2 N(N-1)t}
\left[m_0^2 + \frac{\pi\alpha_s\left<G^2\right>t^2}{6N}\right],
\label{nondiag}\\
&&\Pi_c(t) =
i\frac{C_F N!\,\pi\alpha_s\left<\overline{q}q\right>\widetilde{S}(t)t^3}
{144N^3}.
\nonumber
\end{eqnarray}
The factorization approximation is used for $d=7$ quark-gluon condensates,
so these terms are order-of-magnitude estimate only.
In this approximation the diagram Fig.~\ref{F2}f does not contribute.

Correlators obey the dispersion representation
\begin{equation}
\Pi(\omega) = \int\limits_0^\infty
\frac{\rho(\varepsilon)d\varepsilon}{\varepsilon-\omega-i0} + \cdots,
\quad
\Pi(t) = -\widetilde{S}(t) \int\limits_0^\infty
\rho(\omega) e^{-i\omega t} d\omega + \cdots
\label{disp}
\end{equation}
A subtraction polynomial in $\Pi(\omega)$ (denoted by dots)
leads to $\delta(t)$ and its derivatives in $\Pi(t)$.
We analytically continue correlators from $t>0$ to imaginary $t=-i\tau$.
Then $\Pi(\tau)$ and $\rho(\omega)$ are related by the Laplace transform
\begin{equation}
\Pi(\tau) = i\int\limits_0^\infty \rho(\omega) e^{-\omega\tau} d\omega,
\quad
\rho(\omega) = \frac{1}{2\pi} \int\limits_{a-i\infty}^{a+i\infty}
\Pi(\tau) e^{\omega\tau} d\tau,
\label{laplace}
\end{equation}
where $a$ is to the right from all singularities of $\Pi(\tau)$.
A term $\widetilde{S}(t)/t^n$ ($n\ge1$) in $\Pi(t)$
gives $i^n\vartheta(\omega)\omega^{n-1}/(n-1)!$ in $\rho(\omega)$;
a term $\widetilde{S}(t)t^n$ ($n\ge0$)
corresponds to $(-i)^{n+2}\delta^{(n)}(\omega)$.
We obtain for the diagonal and nondiagonal correlators
\begin{eqnarray}
&&\rho_{\rm d}(\omega) = \frac{N!}{\pi^4}\vartheta(\omega)
\left[\frac{\omega^5}{5!} +
c\frac{\pi\alpha_s\left<G^2\right>\omega}{32N(N-1)}\right]
\nonumber\\
&&\quad{} - \frac{N!\,\left<\overline{q}q\right>^2}{4N^2}
\left[\delta(\omega) - \frac{m_0^2}{8}\delta''(\omega) + \cdots\right],
\label{rho}\\
&&\rho_{\rm n}(\omega) = \frac{N!\,\left<\overline{q}q\right>}{\pi^2 N}
\Bigg[\vartheta(\omega)\frac{\omega^2}{2}
- \tfrac{1}{16}\left(1-\tfrac{c}{N-1}\right)
\left(m_0^2\vartheta(\omega)
+ \frac{\pi\alpha_s\left<G^2\right>}{6N} \delta'(\omega) \right)
\nonumber\\
&&\quad{} - \frac{C_F \pi^3\alpha_s\left<\overline{q}q\right>^3}{144N^2}
\delta'''(\omega) \Bigg].
\nonumber
\end{eqnarray}
We have also verified the leading terms of these formulae
by a direct application of the Cutkosky rules in the momentum space.

We equate the OPE-based expressions~(\ref{diag}--\ref{nondiag}) for $\Pi(\tau)$
to the dispersion representation~(\ref{disp}) with a model spectral density
containing the lowest baryon contribution and the continuum contribution.
This procedure is equivalent
to the nonrelativistic Borel sum rules~\cite{Shuryak}.

Baryon residues in QCD are usually defined as $\left<0|j|B\right>=fu_B$
where the relativistic normalization of the baryon state $\left|B\right>$
and its wave function $u_B$ is implied (and $\Tr\tau^+\tau=1$ is assumed).
This normalization is senseless in HQET;
one should use the nonrelativistic normalization instead.
Both sides of this equation are divided by $\sqrt{2m}$,
and its form does not change.
So baryonic residues $f$ are constant in the heavy quark limit
up to logarithmic renormalization effects.
We have the lowest baryon's contribution to the spectral density
$\rho(\omega)=\frac{f^2}{2}\delta(\omega-\varepsilon)$
where $\varepsilon$ is the HQET baryon's energy
(i.~e.\ mass minus the heavy quark mass).

We adopt the standard continuum model: its spectral density is equal to
the theoretical one~(\ref{rho}) starting at a continuum threshold $\omega_c$
(of course, $\delta^{(n)}(\omega)$ terms don't contribute).
The continuum contribution is thus subtracted from the theoretical part
of the sum rule leaving an integral over the resonance's duality interval
(up to $\omega_c$).
A term $\widetilde{S}(t)/t^n$ after subtracting the continuum contribution
gets the factor $f_{n-1}(\omega_c\tau)$
where $f_n(x)=1-e^{-x}\sum_{m=0}^n\frac{x^m}{m!}$.
Therefore the formulae~(\ref{rho}) are not necessary
to construct the sum rules.
We include all $\widetilde{S}(t)/t^n$ terms with $n\ge1$
to the spectral density, as was done in~\cite{BI}.
If power corrections are not included, it means the absence
of such subtraction for them what leads to their overestimation.

The two diagonal correlators give the identical sum rules
for $\left<0|\widetilde{\jmath}_1|B\right>$
and $\left<0|\widetilde{\jmath}_2|B\right>$.
Therefore these matrix elements are equal with the accepted accuracy.
Following~\cite{Shuryak}, we introduce the dimensionless variables
$\tau=\frac{1}{kE}$,
$f^2=\frac{N!\,\left<\overline{q}q\right>^2}{2N^2}n$,
$m_0=4kE_0$,
$\frac{\pi\alpha_s\left<G^2\right>}{32N(N-1)}=(kE_G)^4$,
$\omega_c=kE_c$,
$\varepsilon=kE_r$,
$k^3=-\frac{\pi^2}{2N}\left<\overline{q}q\right>$.
Then we arrive at the two sum rules
\begin{eqnarray}
&&n e^{-E_r/E} = E^6 \left[ f_5\left(\frac{E_c}{E}\right)
+ c \frac{E_G^4}{E^4} f_1\left(\frac{E_c}{E}\right) \right]
+ \exp\left(-2\frac{E_0^2}{E^2}\right)
\label{sr}\\
&& = 2E^3 \left[ f_2\left(\frac{E_c}{E}\right)
- \left(1-\tfrac{c}{2}\right)\frac{E_0^2}{E^2} f_0\left(\frac{E_c}{E}\right)
+ \tfrac{2}{3}\left(1-{\textstyle\frac{c}{2}}\right)\frac{E_G^4}{E^4}
+ \frac{\alpha_s}{27\pi} \frac{1}{E^6} \right].
\nonumber
\end{eqnarray}
We have used the Gaussian anzatz for the nonlocal quark condensate
in the diagonal sum rule.
The nondiagonal one is used at larger $E$
where the nonlocality is not so important.
For consistency, we use the factorization estimate
for all of the diagrams Fig.~\ref{F2}d--f (including the nonlocality).

Note that $\left<0|\widetilde{\jmath}_1|B\right>$
and $\left<0|\widetilde{\jmath}_2|B\right>$ can't be always equal
because they have different anomalous dimensions;
hence the anomalous dimensions effects are out of our control here.
We obtain these matrix elements normalized at a typical Borel parameter scale;
if one wants to have them at a different normalization point
(e.~g.\ at $m_Q$ for matching with QCD),
one should use the anomalous dimensions~(\ref{adim}).

At the standard value of the quark condensate, the energy unit $k=280$MeV.
It is stated in~\cite{BI} that light baryons are described better
if the quark condensate is reduced by 20\%.
This leads to $k=260$MeV; this value was also used in~\cite{Shuryak}.
We also accept it; the standard values of $m_0^2$ and $\left<G^2\right>$
then give $E_0=0.85$, $E_G=0.60$.

\begin{figure}[p]
\begin{center}
\includegraphics[bb=0 350 290 540]{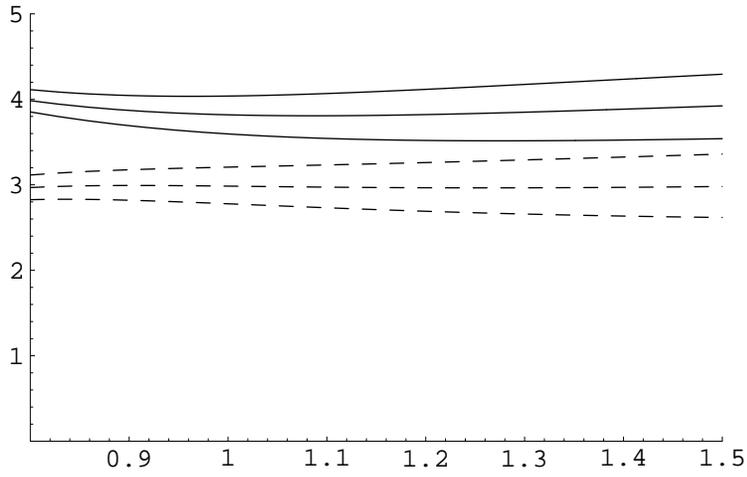}\\
\includegraphics[bb=0 350 290 540]{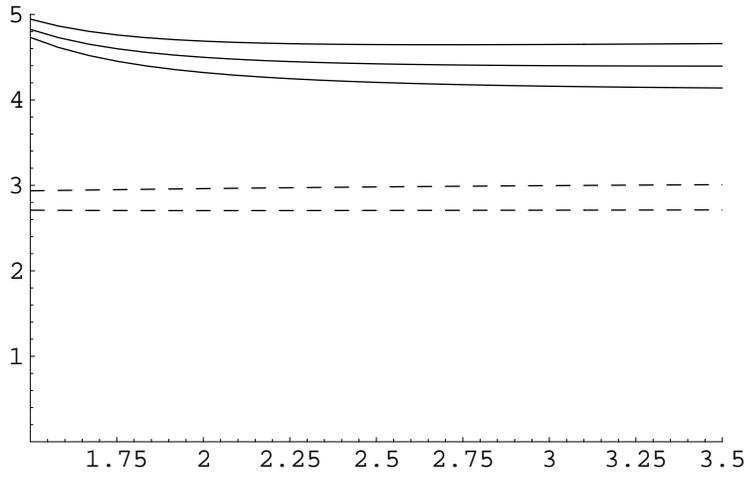}
\end{center}
\caption{The diagonal sum rules.
a) $E_r$ as a function of $E$ at various $E_c$
from the logarithmic derivative of the sum rule:
three lower curves---results for $\Lambda_Q$ at $E_c=4.1$, 4.6, 5.1;
three upper curves---results for $\Sigma^{(*)}_Q$ at $E_c=5.1$, 5.6, 6.1.
b) $n$ as a function of $E$ at various $E_c$, $E_r$:
three lower curves---results for $\Lambda_Q$
at $E_c=4.1$, $E_r=2.75$; $E_c=4.6$, $E_r=2.95$; $E_c=5.1$, $E_r=3.2$;
three upper curves---results for $\Sigma^{(*)}_Q$
at $E_c=5.1$, $E_r=3.6$; $E_c=5.6$, $E_r=3.8$; $E_c=6.1$, $E_r=4.05$.}
\label{F3}
\end{figure}

The results of the diagonal sum rules analyses are shown in Fig.~\ref{F3}.
In the selected range of the Borel parameter $E$,
the perturbative contribution, the gluon condensate one,
and the quark condensate one constitute for $\Lambda_Q$
60--40\%, 25--15\%, and 30--35\%, correspondingly.
For $\Sigma^{(*)}_Q$, the gluon condensate contribution
is 3 times smaller and has the opposite sign.
The continuum contribution
(which is subtracted from the theoretical side of the sum rule)
rapidly grows, and is several times larger than the total result
at the right end of the interval.
This means that the relative error due to the rough continuum model
is multiplied by 1--5 in the total result.
On the other hand, the contribution of poorly known higher power corrections
(with $d\ge8$) becomes large at the left end of the interval.
Namely, the nonlocal quark condensate is several times smaller
than the local one at this point, and the $d=6$ contribution
is almost completely compensated by $d\ge8$ ones.
The remarkable stability of the sum rule in the lower part of this interval
is a plausible argument in favour of the Gaussian anzatz
for the nonlocal quark condensate (the results with, i.~g.,
the factorization anzatz are not so stable in this region).

Thus we obtain
\begin{eqnarray}
\Lambda_Q: &\quad& \varepsilon=780{\rm{}MeV}, \quad
\omega_c=1200{\rm{}MeV}, \quad
f=(1.8{\rm--}2.7)\cdot10^{-2}{\rm{}GeV}^3,
\nonumber\\
\Sigma^{(*)}_Q: &\quad& \varepsilon=990{\rm{}MeV}, \quad
\omega_c=1460{\rm{}MeV}, \quad
f=(2.9{\rm--}4.1)\cdot10^{-2}{\rm{}GeV}^3.
\nonumber
\end{eqnarray}
Taking into account the recent experimental result~\cite{UA1}
$m_{\Lambda_b}=5640\pm50\pm30$MeV, we obtain a reasonable value $m_b=4860$MeV.
The $\Lambda_b$--$\Sigma_b$ splitting of 210MeV is in agreement
with the potential model expectations of 190MeV (see e.~g.~\cite{CI}).
The distances from the resonance energies to the continuum thresholds
in $\Lambda_b$ and $\Sigma_b$ channels are 420MeV and 470MeV;
they approximately correspond to the distances to the first excited states
in these channels expected in the potential model~\cite{CI} 460MeV and 405MeV.

The results of~\cite{Shuryak} for $\Lambda_Q$ are $\varepsilon=700$MeV,
$f=2\cdot10^{-3}$GeV$^3$;
it is an order of magnitude too low\footnote{At the quoted value
of $f_{\Lambda_Q}$, the right-hand side of the equation~(45) in~\cite{Shuryak}
is about two orders of magnitude less than the right-hand side.}.
The sum rule for $\Sigma_Q$ in~\cite{Shuryak} is incorrect.
$\Lambda_c$ and $\Sigma_c$ were considered
in the second paper of~\cite{Chernyak} using relativistic sum rules.
Two matrix elements obtained there reduce
to $\left<0|\widetilde{\jmath}_{\Lambda1}|\Lambda_c\right>$
in the heavy quark limit,
and give $f_{\Lambda_c1}=3.2\cdot10^{-2}$GeV$^3$ and $4\cdot10^{-2}$GeV$^3$;
one matrix element reduces
to $\left<0|\widetilde{\jmath}_{\Sigma2}|\Sigma_c\right>$,
and give $f_{\Sigma_c2}=7.2\cdot10^{-2}$GeV$^3$.

\begin{figure}[p]
\begin{center}
\includegraphics[bb=0 350 290 540]{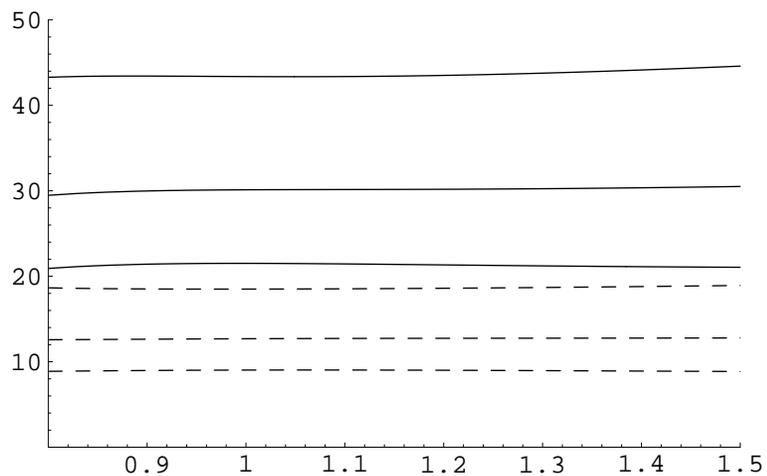}\\
\includegraphics[bb=0 350 290 540]{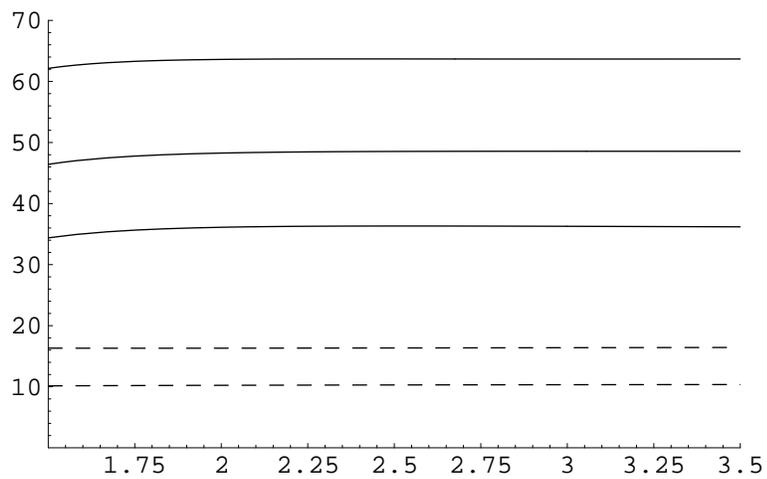}
\end{center}
\caption{The nondiagonal sum rules.
a) $E_r$ as a function of $E$ at various $E_c$
from the logarithmic derivative of the sum rule:
two lower curves---results for $\Lambda_Q$ at $E_c=3.4$, 3.0;
three upper curves---results for $\Sigma^{(*)}_Q$ at $E_c=5.1$, 5.6, 6.1.
b) $n$ as a function of $E$ at various $E_c$, $E_r$:
two lower curves---results for $\Lambda_Q$
at $E_c=3.4$, $E_r=2.65$; $E_c=3.9$, $E_r=2.95$;
three upper curves---results for $\Sigma^{(*)}_Q$
at $E_c=5.1$, $E_r=4.2$; $E_c=5.6$, $E_r=4.4$; $E_c=6.1$, $E_r=4.65$.}
\label{F4}
\end{figure}

The results of the nondiagonal sum rules analyses are shown in Fig.~\ref{F4}.
In the selected range of the Borel parameter $E$,
the $m_0^2$ correction in the $\Lambda_Q$ sum rule reduces starting from 40\%;
in the case of $\Sigma^{(*)}_Q$ it is 7/3 times larger.
Therefore the left end of the interval
is not very reliable for $\Sigma^{(*)}_Q$.
The estimated contribution of poorly known $d\ge7$ corrections
does not exceed 10\%.
The continuum contribution grows not so strongly as in the diagonal case,
allowing us to use larger values of $E$.

The sum rule for $\Lambda_Q$ prefers somewhat lower values
for the effective continuum threshold than in the diagonal case,
and gives the same results for the resonance energy and the residue.
The sum rule for $\Sigma^{(*)}_Q$ gives somewhat too high values
for the mass and the residue.
In general, the agreement of these two completely independent sets
of sum rules gives us more confidence in the results.

\section{Three-point correlators}
\label{Sthree}

Now we shall consider correlators of two HQET baryonic currents
$\widetilde{\jmath}_{1,2}=(q^T C\Gamma_{1,2}q)\Gamma'_{1,2}\widetilde{Q}_{1,2}$
(where $\widetilde{Q}_{1,2}=\widehat{v}_{1,2}\widetilde{Q}_{1,2}$
are the effective heavy quark fields with the velocities $v_{1,2}$,
$v_1\cdot v_2=\ch\varphi$), and a heavy-heavy current
$J=\overline{\widetilde{Q}}_1\Gamma\widetilde{Q}_2$.
They have the structure
\begin{eqnarray}
&&\left<T\widetilde{\jmath}_1(x_1)J(0)\overline{\widetilde{\jmath}}_2(x_2)
\right> =
\left(\Gamma'_1\frac{1+\widehat{v}_1}{2}\Gamma\frac{1+\widehat{v}_2}{2}
\overline{\Gamma}'_2\right) 2\Tr\tau\tau^+
\nonumber\\
&&\quad \int\limits_0^\infty dt_1\delta(x_1-v_1 t_1)
\int\limits_0^\infty dt_2\delta(x_2+v_2 t_2) K(t_1,t_2).
\label{struct3}
\end{eqnarray}
The expression~(\ref{struct3}) without the first factor
is the correlator in which the heavy quark spin is switched off.
For $j^\pi=0^+$ $K(t_1,t_2)$ is a scalar function;
for $j^\pi=1^+$ it is a tensor
$K_{\mu\nu}=K_{||}e_{1||\mu}e_{2||\nu}+K_\bot\delta_{\bot\mu\nu}$,
where $e_{1||}=(v_2-\ch\varphi v_1)/\sh\varphi$,
$e_{2||}=-(v_1-\ch\varphi v_2)/\sh\varphi$
are the light fields' polarization vectors in the scattering plane,
and $\delta_{\bot\mu\nu}=\sum e_{1\bot\mu}e_{2\bot\nu}=
[\ch\varphi(v_{1\mu}v_{2\nu}+v_{2\mu}v_{1\nu})
-v_{1\mu}v_{1\nu}-v_{2\mu}v_{2\nu}]/\sh^2\varphi-g_{\mu\nu}$,
$e_{1\bot}=e_{2\bot}$ are two orthonormalized polarization vectors
orthogonal to this plane.
In each of the three cases ($0^+$, $1^+_{||}$, $1^+_\bot$)
there are two diagonal correlators and one nondiagonal one.

In the limiting case $\varphi=0$ we have
\begin{equation}
K(t_1,t_2) = \Pi(t_1+t_2).
\label{ward}
\end{equation}
Indeed, let's consider any diagram for the two-point correlator
in the coordinate space (for simplicity, with the scalar heavy quark).
Vertices along the heavy quark line have times
$t_0\le t_1\le\cdots\le t_{n-1}\le t_n$,
and integration in $t_1$,\dots, $t_n$ is performed.
Consider the diagrams obtained by inserting the heavy-heavy vertex
(with the time $t$ and $\varphi=0$) to all possible places.
These diagrams have the integration regions
$t_0\le t_1\le\cdots\le t_{m-1}\le t\le t_m\le\cdots\le t_{n-1}\le t_n$.
These regions span the whole integration region of the original diagram,
therefore the sum of all these three-point diagrams
is equal to the two-point diagram.
In the momentum space the equation~(\ref{ward}) reads
$K(\omega_1,\omega_2)=\frac{\Pi(\omega_1)-\Pi(\omega_2)}{\omega_1-\omega_2}$.
For $\Sigma^{(*)}_Q$ correlators at $\varphi=0$ we have $K_{||}=K_\bot$
because all directions orthogonal to $v$ are equivalent.

\begin{figure}[p]
\begin{center}
\includegraphics{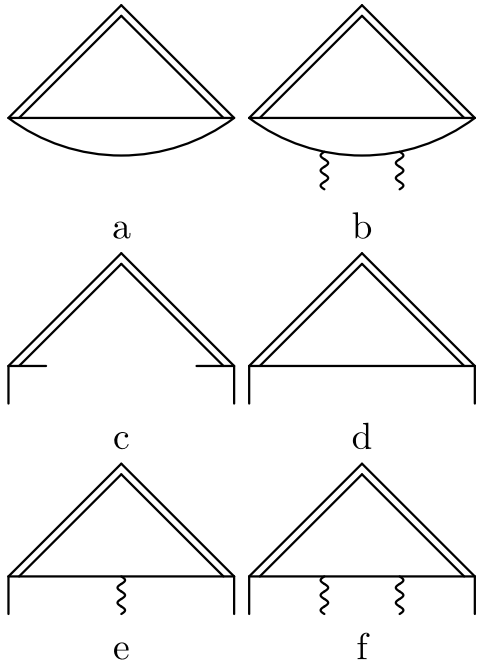}
\end{center}
\caption{Three-point correlator}
\label{F5}
\end{figure}

For the diagonal correlators we obtain (Fig.~\ref{F5})
\begin{eqnarray}
&&K_{a+b}(t_1,t_2) = -i
\frac{N!\,\widetilde{S}(t_1)\widetilde{S}(t_2)a}{\pi^4(x^2)^4}
\Bigg[ 1
\nonumber\\
&&\quad{} + \frac{\pi\alpha_s\left<G^2\right>}{96N(N-1)}
\left(4(N-2)t_1^2 t_2^2 \sh^2\varphi + 3c(x^2)^2\right) \Bigg],
\label{diag3}\\
&&K_c(t_1,t_2) - i \frac{N!\,\widetilde{S}(t_1)\widetilde{S}(t_2)}{4N^2}
\left<\overline{q}(-v_2 t_2)q(v_1 t_1)\right>^2,
\nonumber
\end{eqnarray}
where $x=v_1 t_1+v_2 t_2$, $x^2=t_1^2+t_2^2+2\ch\varphi t_1 t_2$;
$a=x^2$ for $\Lambda1$, $\Sigma_{||}2$, $\Sigma_\bot 1$ cases
(1, 2 refer to $\widetilde{\jmath}_{1,2}$ correlators),
and $a=y^2=\ch\varphi(t_1^2+t_2^2)+2t_1 t_2$
for $\Lambda_2$, $\Sigma_{||}1$, $\Sigma_\bot 2$ cases;
$c=1$ for $\Lambda1$,
$c=\frac13\left(1+2\ch\varphi\frac{x^2}{y^2}\right)$ for $\Lambda2$,
$c=\frac13\left(1-2\ch\varphi\frac{x^2}{y^2}\right)$ for $\Sigma_{||}1$,
$c=-\frac13$ for $\Sigma_{||}2$, $\Sigma_\bot 1$, $\Sigma_\bot 2$.
These results agree with~(\ref{diag}) at $\varphi=0$~(\ref{ward}).

\begin{sloppypar}
The nonlocal quark condensate $\left<\overline{q}(x)q(y)\right>$
in the 0-gauge (i.~e.\ the fixed-point gauge $(x-x_0)_\mu A_\mu(x)=0$
with $x_0=0$) can be rewritten in the gauge-invariant form
$\left<\overline{q}(x)E(x,0)E(0,y)q(y)\right>$,
where $E(x,y)=P\exp(-i)\int_x^y A_\mu(z)dz_\mu$.
Using the translational invariance we rewrite it as
$\left<\overline{q}(0)E(0,-x)E(-x,y-x)q(y-x)\right>$,
or $\left<\overline{q}(0)E(-x,y-x)q(y-x)\right>$ in the new 0-gauge.
In the factorization approximation (giving only a rough estimate)
we can replace $E(x,y)$ in this formula by $\left<\frac{1}{N}\Tr E(x,y)\right>
=1+\frac{\pi\alpha_s\left<G^2\right>}{24N}((xy)^2-x^2 y^2)$.
This gives $\left<\overline{q}(-v_2 t_2)q(v_1 t_1)\right>
=\left<\overline{q}(0)q(x)\right>\left[1+
\frac{\pi\alpha_s\left<G^2\right>t_1^2 t_2^2\sh^2\varphi}{24N}\right]$.
\end{sloppypar}

For the nondiagonal correlators we obtain (Fig.~\ref{F5})
\begin{eqnarray}
&&K_{d+f}(t_1,t_2) = -\frac{N!\,\widetilde{S}(t_1)\widetilde{S}(t_2)
\left<\overline{q}q\right>}{\pi^2 N(x^2)^2}
\nonumber\\
&&\quad\Bigg[ a \left( 1 + \frac{m_0^2 x^2}{16}
+ \frac{\pi\alpha_s\left<G^2\right>}{96N}
\left( (x^2)^2 + 2t_1^2 t_2^2\sh^2\varphi \right) \right)
\nonumber\\
&&\quad{} + b \frac{t_1 t_2\sh^2\varphi}{48}
\left( m_0^2 + \frac{\pi\alpha_s\left<G^2\right>}{3N} \right) \Bigg],
\label{nondiag3}\\
&&K_e(t_1,t_2) = \frac{N!\,\widetilde{S}(t_1)\widetilde{S}(t_2)
\left<\overline{q}q\right>}{16\pi^2 N(N-1)}
\Bigg[ ac \left( \frac{m_0^2}{x^2}
+ \frac{\pi\alpha_s\left<G^2\right>}{6N} \right)
\nonumber\\
&&\quad{} + \frac{2}{3} b \frac{t_1 t_2\sh^2\varphi}{(x^2)^2}
\left( m_0^2 + \frac{\pi\alpha_s\left<G^2\right>}{N}
\left( \frac{a}{b}t_1 t_2 + \frac{x^2}{3} \right) \right) \Bigg],
\nonumber\\
&&K_c(t_1,t_2) = \frac{C_F N!\,\widetilde{S}(t_1)\widetilde{S}(t_2)
  \pi\alpha_s\left<\overline{q}q\right>^3(t_1+\ch\varphi t_2)d}
{144N^3},
\nonumber
\end{eqnarray}
where $a=t_1+\ch\varphi t_2$, $b=t_2$, $d=1$ for $\Lambda$ and $\Sigma_\bot$,
$a=t_2+\ch\varphi t_1$, $b=t_1$, $d=\ch\varphi$ for $\sigma_{||}$;
$c=1$ for $\Lambda$, $c=-\frac13$ for $\Sigma_{||}$, $\Sigma_\bot$.
These results agree with~(\ref{nondiag}) at $\varphi=0$~(\ref{ward}).

\begin{figure}[ht]
\begin{center}
\includegraphics{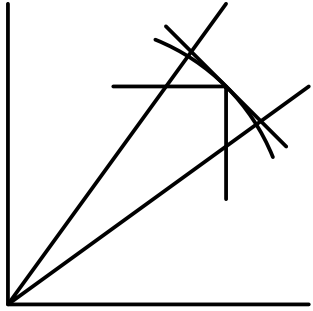}\\
\end{center}
\caption{Physical region of double spectral densities}
\label{F6}
\end{figure}

Three-point correlators obey the double dispersion representation
\begin{eqnarray}
&&K(\omega_1,\omega_2) = \int
\frac{\rho(\varepsilon_1,\varepsilon_2)d\varepsilon_1 d\varepsilon_2}
{(\varepsilon_1-\omega_1-i0)(\varepsilon_2-\omega_2-i0)}
+ \cdots,
\label{disp3}\\
&&K(t_1,t_2) = \widetilde{S}(t_1)\widetilde{S}(t_2)
\int \rho(\omega_1,\omega_2) e^{-i\omega_1 t_1-i\omega_2 t_2}
d\omega_1 d\omega_2 + \cdots
\nonumber
\end{eqnarray}
Subtraction terms in $K(\omega_1,\omega_2)$ (denoted by dots) are polynomials
in $\omega_1$ with coefficients that are arbitrary functions of $\omega_2$
(given by single dispersion integrals), or vice versa.
These terms give $\delta^{(n)}(t_1)$ times arbitrary functions of $t_2$
(or vice versa) in $K(t_1,t_2$.
We analytically continue $K(t_1,t_2)$ from $t_{1,2}>0$
to $t_{1,2}=-i\tau_{1,2}$.
then $K(\tau_1,\tau_2)$ and $\rho(\omega_1,\omega_2)$ are related
by the double Laplace transform
\begin{eqnarray}
&&K(\tau_1,\tau_2) = - \int \rho(\omega_1,\omega_2)
e^{-\omega_1\tau_1-\omega_2\tau_2} d\omega_1 d\omega_2,
\label{laplace3}\\
&&\rho(\omega_1,\omega_2) = \frac{1}{(2\pi)^2}
\int\limits_{a-i\infty}^{a+i\infty} t\tau_1
\int\limits_{a-i\infty}^{a+i\infty} t\tau_2
K(\tau_1,\tau_2) e^{\omega_1\tau_1+\omega_2\tau_2}.
\nonumber
\end{eqnarray}
All considered correlators have the form
$K(\tau_1,\tau_2)=P(\tau_1,\tau_2)/(-x^2)^n$,
where $P(\tau_1,\tau_2)$ is a polynomial,
and $-x^2=\tau_1^2+\tau_2^2+2\ch\varphi\tau_1\tau_2$.
It is convenient to introduce the new variables
$z_1=e^{\varphi/2}\tau_1+e^{-\varphi/2}\tau_2$,
$z_2=e^{\varphi/2}\tau_2+e^{-\varphi/2}\tau_1$,
$\Omega_1=e^{\varphi/2}\omega_1-e^{-varphi/2}\omega_2$,
$\Omega_2=e^{\varphi/2}\omega_2-e^{-varphi/2}\omega_1$.
We have $-x^2=z_1 z_2$,
$\omega_1\tau_1+\omega_2\tau_2=(\Omega_1 z_1+\Omega_2 z_2)/2\sh\varphi$,
and integrals for $\rho(\omega_1,\omega_2)$ factorize.
They don't vanish only in the region $\Omega_1\ge0$, $\Omega_2\ge0$.
This region has the form of a wedge
$e^{-\varphi}\le\frac{\omega_2}{\omega_1}\le e^\varphi$ (Fig.~\ref{F6}).
Therefore it is convenient to use the parameterization
$\omega_{1,2}=\omega(1\pm\eta\th\frac{\varphi}{2})$
(or $\Omega_{1,2}=2\omega\sh\frac{\varphi}{2}(1\pm\eta)$)
in which the physical region is $-1\le\eta\le1$.
It is evident from~(\ref{disp3}) that in the limit $\varphi\to0$
$K(t_1,t_2)$ depends only on $t_1+t_2$ as it should~(\ref{ward}),
and is given by the single dispersion representation~(\ref{disp}) with
\begin{equation}
\rho(\omega) = \lim_{\varphi\to0} 2\omega\th\tfrac{\varphi}{2}
\int\limits_{-1}^{+1} \rho(\omega(1+\eta\th\tfrac{\varphi}{2}),
\omega(1-\eta\th\tfrac{\varphi}{2})) d\eta.
\label{rho2}
\end{equation}

\begin{sloppypar}
For the diagonal correlators we obtain
\begin{equation}
\rho = \frac{N!}{2\pi^4\sh\varphi}
\left[ \frac{\omega^4 A(\eta)}{5!\,\ch^4\frac{\varphi}{2}}
+ \frac{\pi\alpha_s\left<G^2\right>}{32N(N-1)}
\left( \frac{N-2}{6\sh^2\varphi} B(\eta) + C(\eta) \right) \right],
\label{diagr}
\end{equation}
where $A(\eta)=\frac{15}{8}(1-\eta^2)^2$,
$B(\eta) = \delta'(1+\eta)+\delta'(1-\eta)
- \left(3+8\sh^2\frac{\varphi}{2}\right)\allowbreak
\left(\delta(1+\eta)+\delta(1-\eta)-1\right) + 8\sh^4\frac{\varphi}{2}$
for $\Lambda1$, $\Sigma_{||}2$, $\Sigma_\bot 1$;
$A(\eta)=\frac{5}{4}(1-\eta^4)$,
$B(\eta) = \frac{1}{3}\left(\delta''(1+\eta)+\delta''(1-\eta)\right)
- \left(1+4\sh^2\frac{\varphi}{2}\right)
\left(\delta'(1+\eta)+\delta'(1-\eta)\right)
+ 2\left(1+2\sh^2\frac{\varphi}{2}+4\sh^4\frac{\varphi}{2}\right)
\left(\delta(1+\eta)+\delta(1-\eta)-1\right) + 8\sh^4\frac{\varphi}{2}$
for $\Lambda2$, $\Sigma_{||}1$, $\Sigma_\bot 2$;
$C(\eta)=1$ for $\Lambda1$,
$C(\eta)=\frac{1}{3}\left(\delta(1+\eta)+\delta(1-\eta)+2\ch\varphi\right)$
for $\Lambda2$,
$C(\eta)=\frac{1}{3}\left(\delta(1+\eta)+\delta(1-\eta)-2\ch\varphi\right)$
for $\Sigma_{||}1$,
$C(\eta)=-\frac{1}{3}$ for $\Sigma_{||}2$, $\Sigma_\bot 1$,
$C(\eta)=-\frac{1}{3}\left(\delta(1+\eta)+\delta(1-\eta)\right)$
for $\Sigma_\bot 2$.
For the nondiagonal correlators we obtain
\begin{eqnarray}
&&\rho = \frac{\left<\overline{q}q\right>\omega}{2\pi^2\sh\varphi}
\Bigg[ 1 \pm \eta\th\frac{\varphi}{2}
\label{nondiagr}\\
&&\quad{} - \left(1-\tfrac{c}{2}\right)
\frac{m_0^2\ch\frac{\varphi}{2}}{8\omega^2}
\left(e^{\varphi/2}\delta(1\mp\eta)+e^{-\varphi/2}\delta(1\pm\eta)\right)
\Bigg].
\nonumber
\end{eqnarray}
In this formula we assumed $N=3$ because it becomes much simpler in this case.
Upper signs are for $\Lambda$, $\Sigma_\bot$, lower sign---for $\Sigma_{||}$;
$c=1$ for $\Lambda$, $c=-\frac{1}{3}$ for $\Sigma_{||,\bot}$.
Terms with two $\delta$-functions nonvanishing only at the origin
$\omega_1=0$, $\omega_2=0$ are omitted
in equations~(\ref{diagr}--\ref{nondiagr}).
These formulae agree with~(\ref{rho}) at $\varphi=0$~(\ref{rho2}).
We have also verified the leading terms in these formulae
by a direct use of the Cutkosky rules in the momentum space.
\end{sloppypar}

In order to obtain the information on the lowest baryons' form factors,
we equate the OPE-based expressions~(\ref{diag3}--\ref{nondiag3})
for $K(\tau_1,\tau_2)$ to the double dispersion representation~(\ref{disp3})
with a model double spectral density containing
the lowest baryons' contribution in both channels, continuum contribution,
and the mixed contributions with the lowest baryon in the one channel
and continuum in the other one.
The last contribution is exponentially suppressed at sufficiently large
$\tau_{1,2}$, and is neglected as usual~\cite{ISNR}.
The subtraction terms don't contribute at $\tau_1\ne0$, $\tau_2\ne0$.
The applicability regions of the sum rules are symmetric
with respect to the interchange of $\tau_1$ and $\tau_2$
(or nearly symmetric in the nondiagonal case).
Therefore we shall not loose an important information
if we restrict ourselves to the diagonal $\tau_1=\tau_2=\tau/2$.

The contribution of the lowest baryons in both channels to $K(t_1,t_2)$
has the form $\frac{1}{2}f_1 f_2^* e^{-i\varepsilon_1 t_1-i\varepsilon_2 t_2}
\xi(\ch\varphi)$,
where $\xi(\ch\varphi)$ is a scalar function for $j^\pi=0^+$
and $\xi_{\mu\nu}=\xi_{||}e_{1||\mu}e_{2||\nu}+\xi_\bot\delta_{\bot\mu\nu}$
for $j^\pi=1^+$.
The form factors $\xi(\ch\varphi)$ of the $j^\pi=0^+$, $1^+$ baryons
with the heavy quark spin switched off
are called Isgur-Wise~\cite{IW2} functions.
For the form factors of the physical spin symmetry multiplets
$\Lambda_Q$, $\Sigma_Q$, $\Sigma^*_Q$ we have from~(\ref{struct3})
\begin{equation}
\left<B_1|J|B_2\right>
= \overline{u}_1 \Gamma'_1 \Gamma \overline{\Gamma}'_2 u_2 \xi(\ch\varphi).
\label{trace}
\end{equation}
This is equivalent to the results of~\cite{MRR,Falk}.
The contribution to the double spectral density is $\frac12f_1f^*_2
\xi(\ch\varphi)\delta(\omega_1-\varepsilon_1)\delta(\omega_2-\varepsilon_2)$.

We adopt the standard continuum model: the continuum spectral density
is equal to the theoretical one~(\ref{diagr}--\ref{nondiagr})
starting from a smooth curve---continuum threshold.
Note that a non-smooth behavior of the continuum threshold
imposed in the first paper of~\cite{R} has no physical justification,
and leads to an infinite slope of form factors at the origin.
We choose the simplest variant---a straight line continuum threshold
(Fig.~\ref{F6}).
This triangular continuum model works well
in the pion form factor case~\cite{ISNR}.
If the threshold is curved, then the contribution
of the shaded regions in Fig.~\ref{F6} should be subtracted.
this will influence the form factor slope.
This degree of freedom for the slope is analogous
to the freedom of varying the continuum threshold in two-point sum rules,
and should not be very significant if the sum rules are applicable.
With the straight line threshold at $\tau_1=\tau_2$,
we don't need $\rho(\omega_1,\omega_2)$ to write down the sum rule;
the one-variable function $\rho(\omega)$
(given by~(\ref{rho2}) without the limit sign) is enough.
Moreover, this function is proportional to $\omega^n$
where $n$ is evident from the dimensional analysis.
Hence spectral densities~(\ref{diagr}--\ref{nondiagr}) are unnecessary:
we can use terms of the coordinate space results~(\ref{diag3}--\ref{nondiag3})
at $t_1=t_2=-i\tau/2$ multiplied by the corresponding $f_n(\omega_c\tau)$.

Using the dimensionless variables, we obtain the sum rules
\begin{eqnarray}
&&n\xi(\ch\varphi)e^{-E_r/E}
= E^6 \Bigg[ \frac{f_5(E_c/E)}{\ch^6\frac{\varphi}{2}}
\nonumber\\
&&\quad{} + \frac{E_G^4}{E^4} \frac{f_1(E_c/E)}{\ch^2\frac{\varphi}{2}}
\left( c + \tfrac23\sh^2\tfrac{\varphi}{2}
\left( \frac{1}{\ch^2\frac{\varphi}{2}} + \frac{\delta}{2} \right) \right)
\Bigg]
\nonumber\\
&&\quad{} + \exp\left( -2\frac{E_0^2}{E^2}\ch^2\tfrac{\varphi}{2}
+ \tfrac43\frac{E_G^4}{E^4}\sh^2\varphi \right)
\label{sr3}\\
&& = eE^3 \left[ \frac{f_2(E_c/E)}{\ch^2\frac{\varphi}{2}}
- \left(1-\tfrac{c}{2}\right)\frac{E_0^2}{E^2}f_0(E_c/E)
+ \tfrac23\left(1-\tfrac{c}{2}\right)\frac{E_G^4}{E^4}\ch^2\tfrac{\varphi}{2}
\right]
\nonumber\\
&&\quad{} + \frac{\alpha_s d\ch^4\frac{\varphi}{2}}{27\pi E^3}.
\nonumber
\end{eqnarray}
Here $c=1$ for $\Lambda$ and $c=-\frac13$ for $\Sigma$;
in the diagonal sum rule $\delta=1$ for $\Lambda2$,
$\delta=-1$ for $\Sigma_{||}1$, and $\delta=0$ otherwise;
in the nondiagonal sum rule $d=\ch\varphi$ for $\Sigma_{||}$
and $d=1$ otherwise.
In the diagonal sum rule we have included the effect of noncollinearity
in the nonlocal quark condensate to the exponent;
this is an order-of-magnitude estimate only,
and should not be used when this effect is significant.
At $\varphi=0$ these sum rules coincide with~(\ref{sr}).
Dividing~(\ref{sr3}) by~(\ref{sr}), we obtain the formulae
for the Isgur-Wise form factors not containing $E_r$;
the normalization at $\varphi=0$ is automatically correct.

\begin{figure}[p]
\begin{center}
\includegraphics[bb=0 350 290 540]{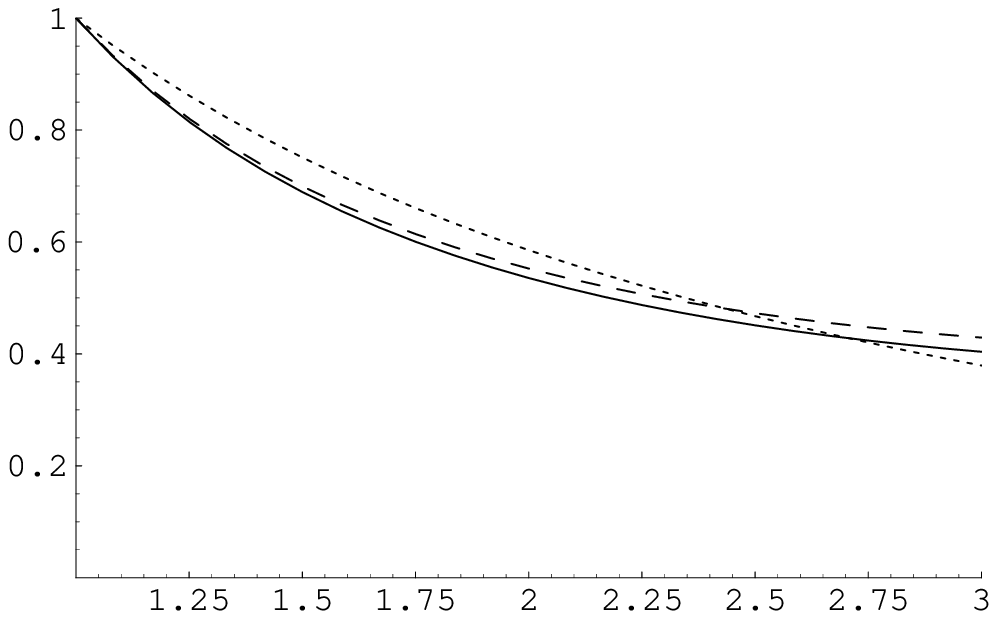}
\includegraphics[bb=0 350 290 540]{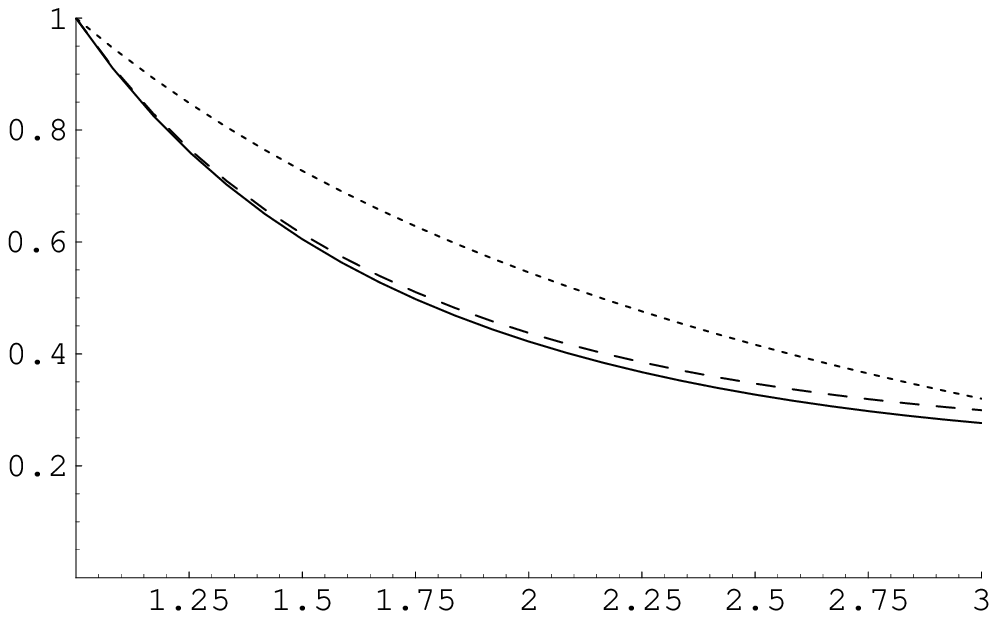}
\end{center}
\caption{Sum rules for the Isgur-Wise form factors.
a) $\Lambda_Q$:
lower curve---the diagonal sum rule 1 at $E_c=4.6$, $E=1.2$;
middle curve---the diagonal sum rule 2 at the same values;
upper curve---the nondiagonal sum rule at $E_c=3.9$, $E=3$.
b) $\Sigma^{(*)}_Q$:
lower curve---the diagonal sum rule 1 for $\Sigma_{||}$ at $E_c=5.6$, $E=1.2$;
middle curve---the diagonal sum rule 2 for $\Sigma_{||}$
and both diagonal sum rules for $\Sigma_\bot$ at the same values;
upper curve---the nondiagonal sum rule for $\Sigma_{||}$ at $E_c=5.6$, $E=3$
(for $\Sigma_\bot$ it differs by less than the line width).}
\label{F7}
\end{figure}

The results of the sum rules analyses for $\xi(\ch\varphi)$
are shown in Fig.~\ref{F7}.
We have taken the best values of the continuum threshold $E_c$
and the intervals of the Borel parameter $E$ from the two-point sum rules.
Variation of these parameters in reasonable bounds
leads only to slight changes of these curves.
The diagonal sum rules are very stable with respect to varying $E$,
but only at $E\ge1.2$; this bound is higher than in the two-point case.
This can be explained by a not very accurate consideration
of the noncollinearity effect in the nonlocal quark condensate.
In the case of $\Lambda_Q$, the two diagonal sum rules
are in a reasonable agreement with each other;
the nondiagonal sum rule predicts somewhat lower slope and a more straight
shape, but is not in a sharp disagreement with the diagonal ones.
In the case of $\Sigma^{(*)}_Q$, the second diagonal sum rule predicts
the equal form factors for the two possible light fields' polarizations;
the first one yields the same result for $\Sigma_\bot$,
but a somewhat different result for $\Sigma_{||}$.
Therefore the accuracy of the approach doesn't allow us to distinguish
between $\xi_{||}(\ch\varphi)$ and $\xi_\bot(\ch\varphi)$.
The results of the diagonal sum rules are in a reasonable agreement
with each other.

On the other hand, the nondiagonal sum rule predicts
a significantly lower slope and a more straight shape.
the nondiagonal sum rules for $\Sigma_{||}$ and $\Sigma_\bot$
differ only in the very small $d=9$ $\left<\overline{q}q\right>^3$ term,
and the predicted curves are indistinguishable.
We remind that in the two-point case an agreement between
the nondiagonal sum rule and the diagonal ones for $\Sigma^{(*)}_Q$
also was much worse than for $\Lambda_Q$.
It would be interesting to understand the reason of this poor agreement.

Note that in the case of the nucleon form factors
a thorough analysis of the sum rules appeared impossible,
and a simplified local duality approach was used~\cite{NR}.
It corresponds to working at infinite Borel parameters,
and using the continuum threshold from the two-point sum rules.
Moreover, only one correlator was used,
so no self-consistency check was possible.
Here we are in a somewhat better position:
we do have wide stability regions in Borel parameters,
and comparison of different correlators allows us to estimate the accuracy
(though it is not high).

{\bf Acknowledgments.} We are grateful to D.~J.~Broadhurst, V.~L.~Cher\-nyak,
and V.~S.~Fadin for fruitful discussions.
A.~G.~G.\ thanks P.~Ball, J.~G.~K\"orner, and K.~Schilcher
for stimulating discussions.


\begin{thebibliography}{99}

\raggedright

\newcommand{\PLB}[3]{Phys.\ Lett.\ {\bf B#1} (19#2) #3}
\newcommand{\NPB}[3]{Nucl.\ Phys.\ {\bf B#1} (19#2) #3}
\newcommand{\PRD}[3]{Phys.\ Rev.\ {\bf D#1} (19#2) #3}
\newcommand{\PRL}[3]{Phys.\ Rev.\ Lett.\ {\bf #1} (19#2) #3}
\newcommand{\ZPC}[3]{Zeit.\ Phys.\ {\bf C#1} (19#2) #3}
\newcommand{\SJNP}[3]{Sov.\ J.\ Nucl.\ Phys.\ {\bf #1} (19#2) #3}
\newcommand{\IJMPA}[3]{Int.\ J.\ Mod.\ Phys.\ {\bf A#1} (19#2) #3}
\newcommand{\FP}[3]{Fortschr.\ Phys.\ {\bf #1} (19#2) #3}
\newcommand{\Prep}[3]{Preprint #1 (19#2)}

\bibitem{Shuryak}
E.~V.~Shuryak. \NPB{198}{82}{83}.

\bibitem{EH}
E.~Eichten, B.~Hill. \PLB{234}{90}{511}.

\bibitem{GG}
B.~Grinstein. \NPB{339}{90}{253};
H.~Georgi. \PLB{240}{90}{447}.

\bibitem{IW}
N.~Isgur, M.~B.~Wise. \PLB{232}{89}{113}; {\bf B237} (1990) 527.

\bibitem{FGGW}
A.~F.~Falk, H.~Georgi, B.~Grinstein, M.~B.~Wise. \NPB{343}{90}{1}.

\bibitem{IW2}
N.~Isgur, M.~B.~Wise. \NPB{348}{91}{276}.

\bibitem{MRR}
H.~Georgi. \NPB{348}{91}{293};
T.~Mannel, W.~Roberts, Z.~Ryzak. \NPB{355}{91}{38}.

\bibitem{Hussain}
F.~Hussain, J.~G.~K\"orner, M.~Kr\"amer, G.~Thompson. \ZPC{51}{91}{321};
F.~Hussain, D.~Liu, M.~Kr\"amer, J.~G.~K\"orner, S.~Tawfiq. \NPB{370}{92}{259}.

\bibitem{Falk}
A.~F.~Falk. \Prep{SLAC-PUB-5689, Stanford}{91}.

\bibitem{super}
H.~Georgi, M.~B.~Wise. \PLB{243}{90}{279};
C.~D.~Carone. \PLB{253}{91}{408}.

\bibitem{SVZ}
M.~A.~Shifman, A.~I.~Vainshtein, V.~I.~Zakharov. \NPB{147}{79}{385}; 448; 519.

\bibitem{SV}
M.~B.~Voloshin, M.~A.~Shifman. \SJNP{45}{87}{292};
H.~D.~Politzer, M.~B.~Wise. \PLB{206}{88}{681}; {\bf B208}{88}{504}.

\bibitem{BG}
D.~J.~Broadhurst, A.~G.~Grozin. \PLB{267}{91}{105};
X.~Ji, M.~J.~Musolf. \PLB{257}{91}{409}.

\bibitem{BG2}
D.~J.~Broadhurst, A.~G.~Grozin. \PLB{274}{92}{421};
E.~Bagan, P.~Ball, V.~M.~Braun, H.~G.~Dosh.
\Prep{HD-THEP-91-36, Heidelberg}{91}.

\bibitem{R}
A.~V.~Radyushkin. \PLB{271}{91}{218};
M.~Neubert. \Prep{SLAC-PUB-5712, Stanford}{91};
D.~J.~Broadhurst, A.~G.~Grozin, O.~I.~Yakovlev. In preparation.

\bibitem{ISNR}
B.~L.~Ioffe, A.~V.~Smilga. \PLB{114}{82}{353}; \NPB{216}{83}{373};
V.~A.~Nesterenko, A.~V.~Radyushkin. \PLB{115}{82}{410}.

\bibitem{Ioffe}
B.~L.~Ioffe. \NPB{188}{81}{175}; Erratum: {\bf B191} (1981) 591.

\bibitem{BI}
V.~M.~Belyaev, B.~L.~Ioffe. JETP {\bf 83} (1982) 876.

\bibitem{CDKS}
Y.~Chung, H.~G.~Dosh, M.~Kremer, D.~Schall.
\PLB{102}{81}{175}; \NPB{197}{82}{55}; \ZPC{15}{82}{367}; {\bf C25} (1984) 151;
D.~Espriu, P.~Pascual, R.~Tarrach. \NPB{214}{83}{285}.

\bibitem{Chernyak}
V.~L.~Chernyak, I.~R.~Zhitnitsky. \NPB{246}{84}{52}; {\bf B345} (1990) 137;
V.~L.~Chernyak, A.~A.~Ogloblin, I.~R.~Zhitnitsky. \ZPC{42}{89}{569}.

\bibitem{Peskin}
M.~E.~Peskin. \PLB{88}{79}{128}.

\bibitem{PS}
A.~A.~Pivovarov, L.~R.~Surguladze. \SJNP{48}{88}{1856};
A.~A.~Ovchinnikov, A.~A.~Pivovarov, L.~R.~Surguladze. \IJMPA{6}{91}{2025}.

\bibitem{NR}
V.~A.~Nesterenko, A.~V.~Radyushkin. \PLB{128}{83}{439}.

\bibitem{TechRev}
V.~A.~Novikov, M.~A.~Shifman, A.~I.~Vainshtein, V.~I.~Zakharov.
\FP{32}{84}{585};
A.~G.~Grozin. \Prep{INP 92-11, Novosibirsk}{92}.

\bibitem{MR}
S.~V.~Mikhailov, A.~V.~Radyushkin. \SJNP{49}{89}{494}.

\bibitem{UA1}
C.~Albajar et al.\ (UA1 collaboration). \Prep{CERN PPE/91-202}{91}.

\bibitem{CI}
S.~Capstick, N.~Isgur. \PRD{34}{86}{2809}.

\end{thebibliography}
\end{document}